\documentclass[pdftex,letterpaper,rmp,twocolumn,groupedaddress,floatfix]{revtex4}
\bibpunct{(}{)}{,}{n}{,}{,}
\usepackage[utf8]{inputenc}
\usepackage{graphicx}
\usepackage{amssymb}
\usepackage{fancybox}
\usepackage{amsmath}
\usepackage{color}
\usepackage{rotating}
\definecolor{darkgray}{rgb}{0.25,0.25,0.25}
\definecolor{darkred}{rgb}{0.89,0.10,0.11}
\definecolor{darkblue}{rgb}{0.12,0.39,0.62}
\usepackage{url}
\usepackage[pdftex,breaklinks=true,colorlinks=true,citecolor=black,linkcolor=black,menucolor=black,urlcolor=darkblue,pdfborder={1 0 0}]{hyperref}
\hypersetup{pdftitle={Mapping change in large networks},pdfauthor={Martin Rosvall and Carl T. Bergstrom, 2009}}

\begin{document}
\makeatletter
\renewcommand\@biblabel[1]{#1.}
\makeatother
\title{Mapping change in large networks}
\author{M. Rosvall}
\email{rosvall@u.washington.edu}
\homepage{http://www.tp.umu.se/~rosvall}
\affiliation{Department of Biology, University of Washington, Seattle, Washington, United States of America}
\author{C. T. Bergstrom}
% \email{cbergst@u.washington.edu}
\affiliation{Department of Biology, University of Washington, Seattle, Washington, United States of America}
\affiliation{Santa Fe Institute, Santa Fe, New Mexico, United States of America}
% \homepage{http://octavia.zoology.washington.edu/}
\date{\today}

\begin{abstract}
Change is a fundamental ingredient of interaction patterns in biology, technology, the economy, and science itself: Interactions within and between organisms change; transportation patterns by air, land, and sea all change; the global financial flow changes; and the frontiers of scientific research change. Networks and clustering methods have become important tools to comprehend instances of these large-scale structures, but without methods to distinguish between real trends and noisy data, these approaches are not useful for studying how networks change. Only if we can assign significance to the partitioning of single networks can we distinguish meaningful structural changes from random fluctuations. Here we show that bootstrap resampling accompanied by significance clustering provides a solution to this problem. To connect changing structures with the changing function of networks, we highlight and summarize the significant structural changes with alluvial diagrams and realize de Solla Price's vision of mapping change in science: studying the citation pattern between about 7000 scientific journals over the past decade, we find that neuroscience has transformed from an interdisciplinary specialty to a mature and stand-alone discipline.

\end{abstract}

\maketitle

\section*{Introduction}
Researchers have developed a suite of network mapping tools to highlight important features while simplifying the overall structure of social and biological systems \cite{girvan_newman,palla,palla2007,guimera-nature,RosvallBergstrom08,fortunato}. With such tools we can abstract, quantify, and comprehend the nature of systems with numerous and diverse interacting components. As powerful as these tools have proven to be for understanding a system's structure, we do not yet have an adequate tool for mapping how this structure {\em changes}. For example: How has the network of global air traffic changed over the past half century? How does the organization of social contacts change when diseases develop and spread? How does the network structure of the federal funds market change when credit markets freeze up? How do gene regulatory networks differ between cancer and non-cancer states? And how does science itself evolve as research tools, strategies, and agendas shift through time? To quantify change in large networks, we must first identify the important structures, then assess which of these structures are statistically significant, and finally capture how these structures change.

Any tool for analyzing change must distinguish between meaningful trends and statistical noise. For example, statistical network models and stratified data make it possible to estimate global properties from the observation of sample networks \cite{thompson2002,hanneman05,handcock2008}. But when we are interested in the unique identities of the individual network components --- Chicago O'Hare plays a unique and irreplaceable role in the global air traffic network, for example --- we need another approach. Recent network approaches have become prominent in the study of complex systems because they can capture and respect the identities and characteristics of the components \cite{albert2002,newmanSIAM}. Often these individual differences matter critically and clustering rather than stratification must be employed to comprehend the data \cite{girvan_newman,palla,palla2007,guimera-nature,RosvallBergstrom08,fortunato}.

Moreover, many of the systems to which we apply network approaches are idiosyncratic in nature and preclude replicate observations. There is one and only one global air traffic network for the year 2009, for example. Therefore we cannot establish statistical significance by looking at multiple samples. Nor can we rely on temporal stability. While structures that remain unchanging over time may be statistically significant, we will not find significant changes by looking for features that stay the same.

One possibility would be to use a resampling technique such as the bootstrap, which assesses the accuracy of an estimate by resampling from the empirical distribution of observations \cite{efron1993}. But we have only a single observation, a single network --- so from what can we resample? When the single observation is composed of numerous components, as a network is composed of nodes and links, we can use the parametric bootstrap to assemble bootstrap networks by resampling from the components. Instead of resampling directly from the empirical distribution, a parametric model is used to fit the data. For the networks discussed in this paper, resampling nodes is not the right approach --- it makes no sense to talk about the US air transit network without O'Hare, let alone the US network with two O'Hares. However, the link weights can be parametrized and resampled without undermining the individual characteristics of the nodes.
With this approach we can assess the significance of clusters and estimate the accuracy of summary statistics, based on the proportion of bootstrap networks that support the observation (see Fig.~\ref{significanceclustering}).

Finally, to reveal stories in the network data and to be able to connect structural and functional changes, we use {\em alluvial diagrams} to highlight and summarize the significant structural changes. Our method could be applied to study, for example, how the global flight traffic pattern changes over time or how the federal funds market adapts structurally to cope with disturbances, but here we illustrate the method by mapping change in the structure of science itself \cite{price}.

\section*{Results}

\subsection*{Journal Citation Networks}
Science is a dynamic, organized, and massively parallel human endeavor to discover, explain, and predict the nature of the physical world. 
In science, new ideas are built upon old ideas. Through cumulative cycles of modeling and experimentation, scientific research undergoes constant change: scientists self-organize into fields that grow and shrink, merge and split. Citation patterns among scientific journals allow us to track this flow of ideas and how the flow of ideas changes over time \cite{price}. Here we use citation data from Thomson-Reuters' Journal Citation Reports 1997--2007, which aggregate, at the journal level, approximately 35,000,000 citations from more than 7000 journals over the past decade. We include citations from articles published in a given year to articles published in the previous two years and, because we are interested in relationships between journals, we exclude journal self-citations.

\subsection*{Significance Clustering}
\begin{figure}[tbp]
\centering
\includegraphics[width=0.85\columnwidth]{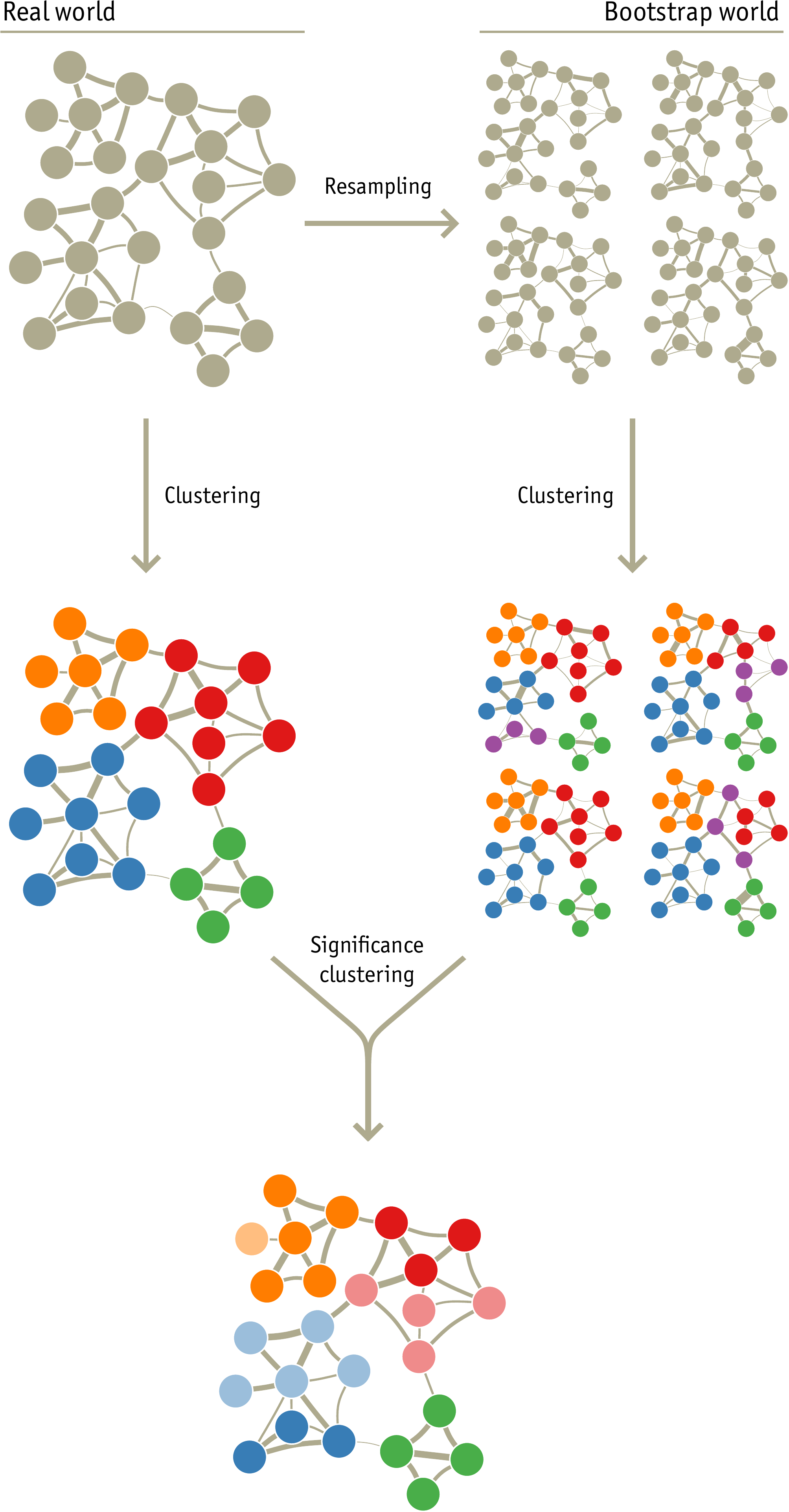}
\caption{\label{significanceclustering}
Significance clustering of networks. The standard approach to cluster networks is to minimize an objective function over possible partitions of the network, as in the left side of the diagram. By repeated resampling of the weighted links from the original network, we create a ``bootstrap world'' of resampled networks. By clustering these as well, and comparing to the clustering of the original network, we can estimate the degree of support that the data provide in assigning each node to a cluster. In the bottom network, the darker nodes are clustered together in at least 95\% of the 1000 bootstrap networks.}
\end{figure}

We first cluster the networks with the information-theoretic clustering method presented in ref.~\cite{RosvallBergstrom08}, which can reveal regularities of information flow across directed and weighted networks. We emphasize that, with appropriate modifications, our method of bootstrap resampling accompanied by significance clustering is general and works for any type of network and any clustering algorithm (see Materials and Methods for a detailed description of the method). To assess the accuracy of a clustering, we resample a large number $B \approx 1000$ of bootstrap networks from the original network. For the directed and weighted citation network of science, in which journals correspond to nodes and citations to directed and weighted links, we treat the citations as independent events and resample the weight of each link from a Poisson distribution with the link weight in the original network as mean. This \emph{parametric} resampling of citations approximates a \emph{non-parametric} resampling of articles, which makes no assumption about the underlying distribution (see, for example, refs.~\cite{karrer_newman,gfeller,costenbader2003} for other resampling techniques). Figure \ref{significanceclustering} illustrates an example network, the clustering of this network, and the clusterings of four of its bootstrap networks. For scalar summary statistics, it is straightforward to assign a 95\% bootstrap confidence interval as spanning the 2.5th and 97.5th percentiles of the bootstrap distribution \cite{costenbader2003}, but working with sets and assessing the accuracy of the clusters requires a different approach.

\begin{figure}
\centering
\includegraphics[width=0.85\columnwidth]{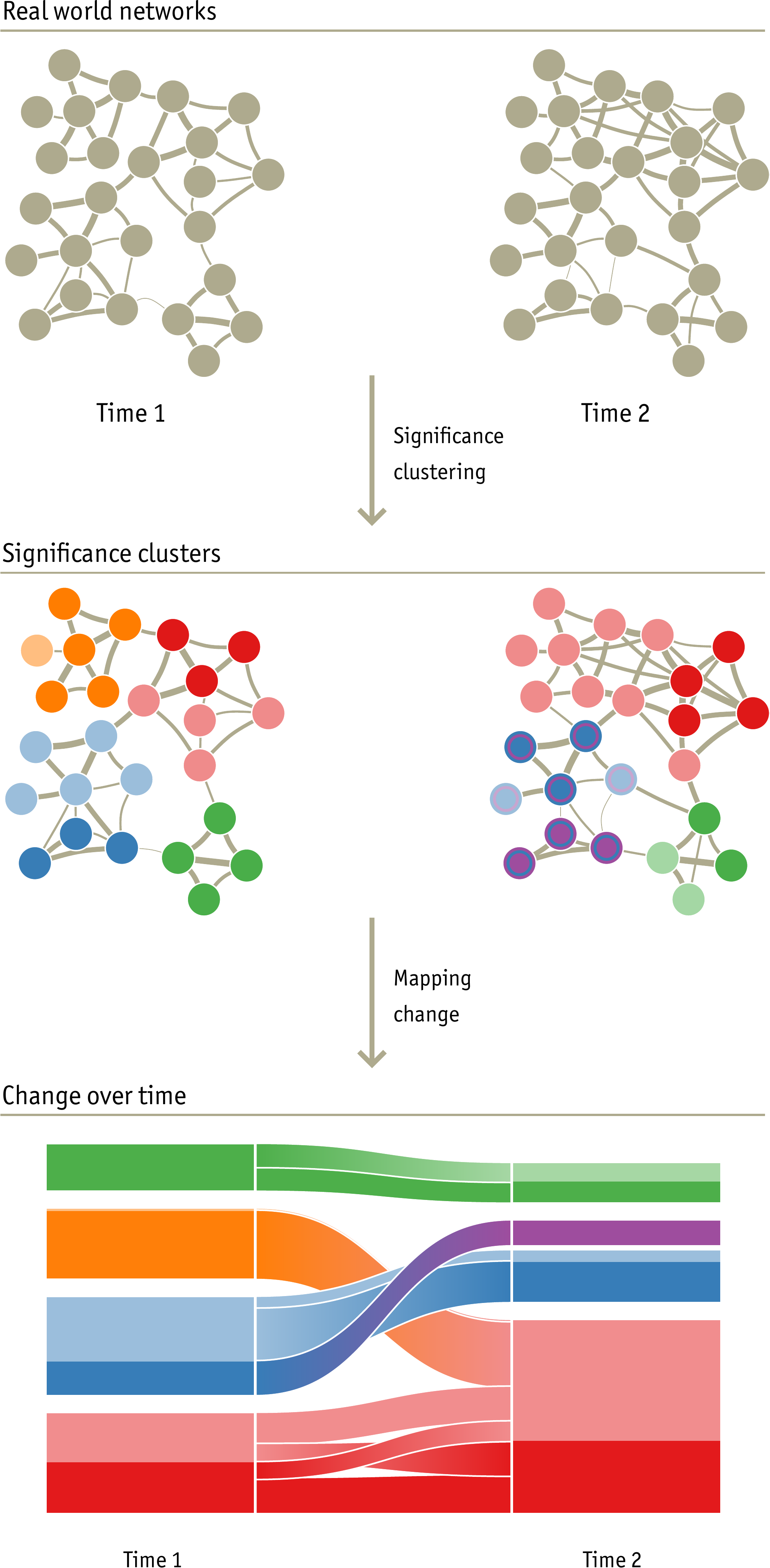}
\caption{\label{networktosankey}
Mapping change in networks. An alluvial diagram (bottom), with clusters ordered by size, reveals changes in network structures over time. 
Here the height of each block represents the volume of flow through the cluster, with significant subsets in darker color.
The orange module merges with the red module, but the nodes are not clustered together in 95\% of the bootstrap networks. The blue module splits, but the significant nodes in the blue and purple modules are clustered together in more than 5\% of the bootstrap networks. With a 5\% significance threshold, neither change is significant.}
\end{figure}

To identify the journals that are significantly associated with the clusters to which they are assigned, we use simulated annealing to search for the largest subset of journals within each cluster of the original network that are clustered together in at least 95\% of all bootstrap networks. To identify the clusters that are significantly distinct from all other clusters, we search for clusters whose significant subset is clustered with no other cluster's significant subset in at least 95\% of all bootstrap networks (see Materials and Methods).  The significance-clustering step of Fig.~\ref{networktosankey} illustrates this process as applied to a network at two different time points. 

\subsection*{Alluvial Diagrams}
Once we have a significance cluster for the network at each time point (or each state), we want to reveal the trends in our data:  we need to simplify and highlight the structural changes between clusters. In the mapping-change step of Fig.~\ref{networktosankey}, we show how to construct an  \emph{alluvial diagram} of the example networks that highlights and summarizes the structural differences between the time 1 and time 2 significance clusters. Each cluster in the network is represented by an equivalently colored block in the alluvial diagram. Darker colors represent nodes that are assigned with statistical significance, while lighter colors represent non-significant assignments. Changes in the clustering structure from one time period to the next are represented by the mergers and divergences that occur in the ribbons linking the blocks at time 1 and time 2.

\begin{figure*}[tbp]
\centering
\includegraphics[width=1.0\textwidth]{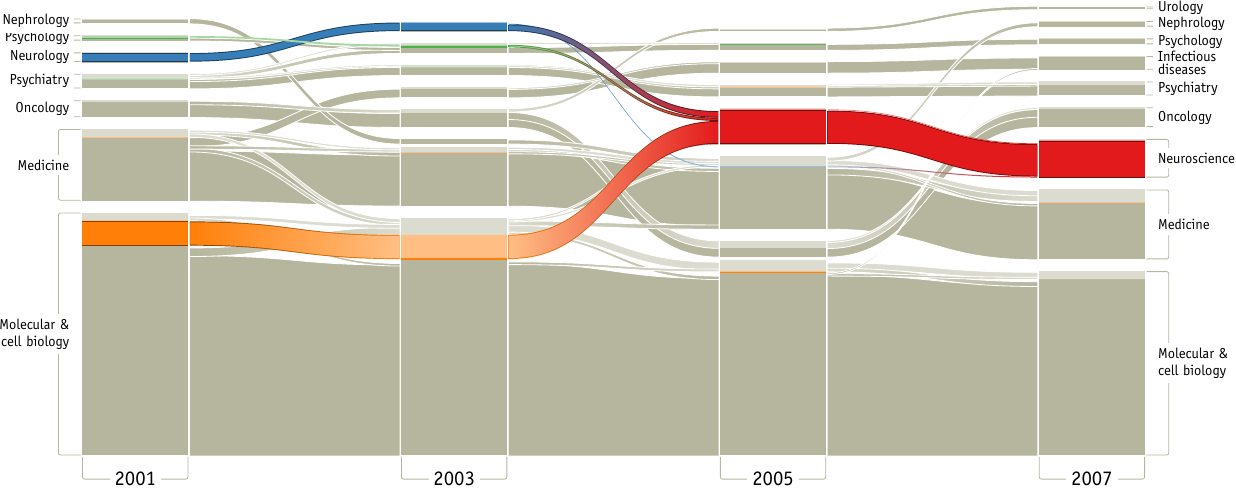}
\caption{\label{neurosankey} Mapping change in science.
This set of scientific fields show the major shifts in the last decade of science. Each significance clustering for the citation networks in years 2001, 2003, 2005, and 2007 occupies a column in the diagram and is horizontally connected to preceding and succeeding significance clusterings by stream fields. Each block in a column represents a field and the height of the block reflects citation flow through the field. The fields are ordered from bottom to top by their size with mutually nonsignificant fields placed together and separated by half the standard spacing. We use a darker color to indicate the significant subset of each cluster. All journals that are clustered in the field of neuroscience in year 2007 are colored to highlight the fusion and formation of neuroscience.}
\end{figure*}

The alluvial diagram for the citation data reveals the significant structural changes that have occurred in science over the past decade.
Rather than viewing the entire diagram, let us highlight a couple of interesting stories. Figure \ref{neurosankey} shows a subset of biomedical fields for the years 2001, 2003, 2005, and 2007. 

The alluvial diagram illustrates, for example, how over the years 2001--2005, urology gradually splits off from oncology and how the field of infectious diseases becomes a unique discipline, instead of a subset of medicine, in 2003. But these changes are just two of many over this period. In the same diagram, we also highlight the biggest structural change in scientific citation patterns over the past decade: the transformation of neuroscience from interdisciplinary specialty to a mature and stand-alone discipline, comparable to physics or chemistry, economics or law, molecular biology or medicine. In 2001, 102 neuroscience journals, lead by \emph{the Journal of Neuroscience}, \emph{Neuron}, and \emph{Nature Neuroscience}, are assigned with statistical significance to the field of molecular and cell biology (dark orange, 84 of 102 journals are assigned significantly). Further, \emph{Brain, Behavior, and Immunity}, \emph{Journal of Geriatric Psychiatry and Neurology}, \emph{Psychophysiology}, and 33 other journals appear with statistical insignificance in psychology (green, 6 of 36 journals are assigned significantly) and \emph{Neurology}, \emph{Annals of Neurology}, \emph{Stroke} and 77 other journals appear with statistical significance in neurology (blue, 75 of 80 journals are assigned significantly).  In 2003, many of these journals remain in molecular and cell biology, but their assignment to this field is no longer significant (light orange, 5 of 102 journals are assigned significantly).  The transformation is underway. In 2005, neuroscience first emerges as an independent discipline (red). The journals from molecular biology split off completely from their former field and have merged with neurology and a subset of psychology into the significantly stand-alone field of neuroscience.

In their citation behavior, neuroscientists have finally cleaved from their traditional disciplines and united to form what is now the fifth largest field in the sciences (after molecular and cell biology, physics, chemistry, and medicine). Although this interdisciplinary integration has been ongoing since the 1950s \cite{cowan2000}, only in the last decade has this change come to dominate the citation structure of the field and overwhelm the intellectual ties along traditional departmental lines. 

\section*{Discussion}
The problem of detecting structural change in large networks adds two new challenges to the basic problem of network clustering: (1) we need appropriate statistical methods to identify significant features of network clustering and to distinguish between trends and noise in the data, and (2) we require effective visualizations to bring out the stories implicit in a time series of cluster maps. To resolve the first of these challenges, we have developed a method for significance clustering based on the parametric bootstrap. To address the second, we have presented the visualization technique of alluvial diagrams. These methods are general to many types of networks and can answer questions about structural change in science, economics, and business.

\section*{Materials and Methods}
Here we lay out the details of how we generate \emph{significance clusters} and \emph{alluvial diagrams} for mapping change in networks.
Because this method assesses how much confidence we should have in the clustering of a network, we can detect, highlight, and simplify the significant structural changes that occur over time or between states in large networks, for example, citation networks, traffic networks, and monetary flow networks.
This approach to mapping change in large networks works for any clustering algorithm. The choice of algorithm depends on the network type (undirected, directed, unweighted, weighted) and the scope of the study. Here we focus on the general case of weighted directed networks. We also assume that the weight of the links can be described by a Poisson-like process. That is, the weights represent, or can be modeled by, independent events in time. This can be generalized to other distributions of link weights; see section 2 below.

\begin{figure*}[ht]
\centering
\includegraphics[width=0.85\textwidth]{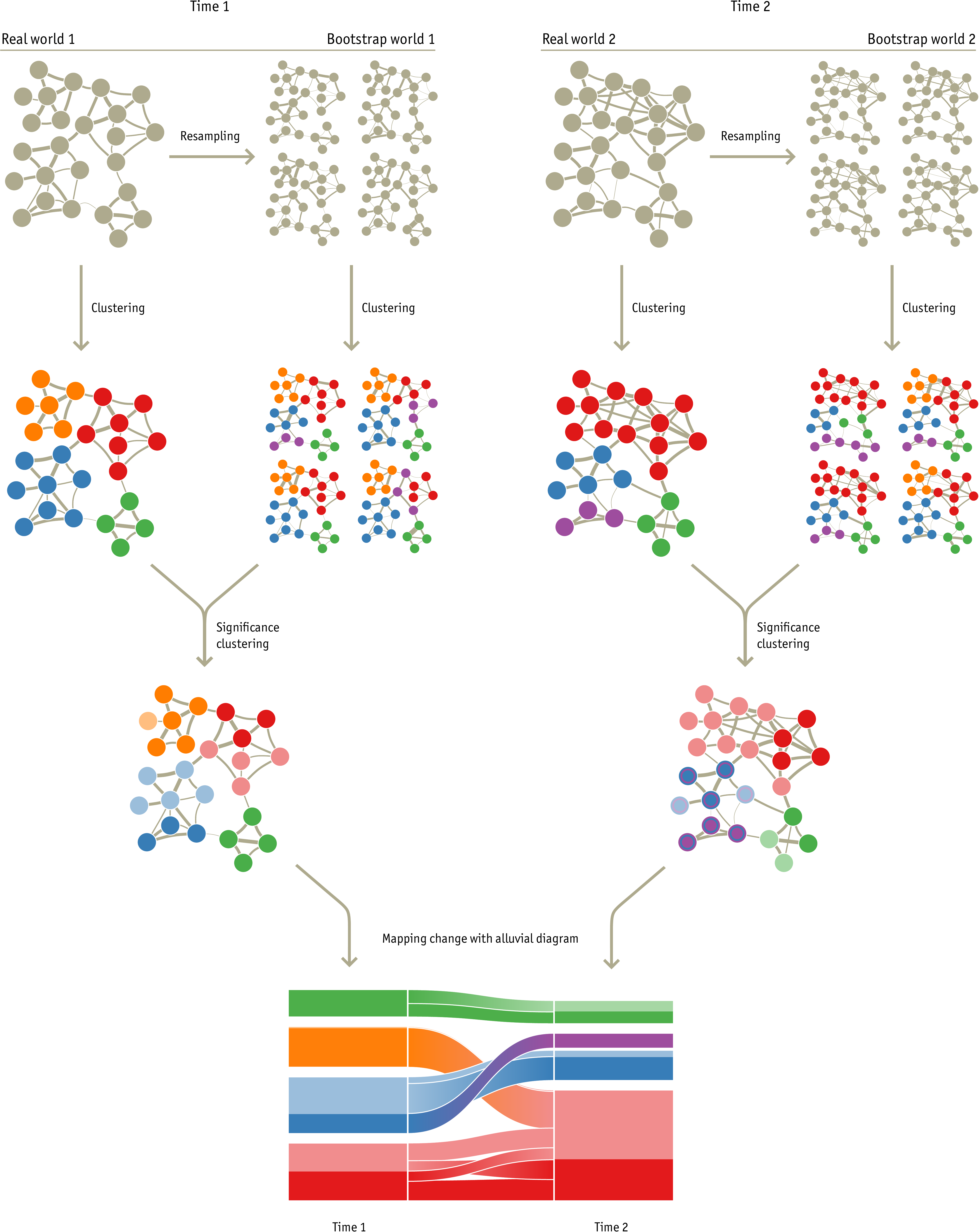}
\caption{Significance clustering and alluvial diagram for mapping change in large networks. By repeatedly resampling of the weighted links from the original networks, we create ``bootstrap worlds'' of 1000 resampled networks. By clustering these bootstrap networks, and comparing to the clustering of the original networks, we can estimate the degree of support that the data provide in assigning each node to a cluster. In the bottom networks, the darker colors represent nodes that are clustered together in at least 95\% of the 1000 bootstrap networks. The alluvial diagram highlights and summarizes the structural changes between the time 1 and time 2 significance clusters. The height of each block represents the volume of flow through the cluster. The clusters are ordered from bottom to top by their size, with mutually nonsignificant clusters placed together and separated by a third of the standard spacing. The orange module merges with the red module, but the nodes are not clustered together in 95\% of the bootstrap networks. The blue module splits, but the significant nodes in the blue and purple modules are clustered together in more than 5\% of the bootstrap networks. Neither change is significant.}
\label{networktosankey_SI}
\end{figure*}

The method consists of four steps, described below and illustrated in  Fig.~\ref{networktosankey_SI}:
\begin{enumerate}
\item Cluster the original networks observed at each time point.
\item Generate and cluster the bootstrap replicate networks for each time point.
\item Determine significance of the clustering for at each time point.
\item Generate an alluvial diagram to illustrate changes between time points.
\end{enumerate}
For simplicity of description, here we map the change between two states $G^1$ and $G^2$ of a network --- but it is straightforward to extend the procedure to more states. We enumerate the $N$ nodes by $\alpha = 1,2,\ldots,N$. (The set of nodes in $G^1$ need not be identical to the set in $G^2$.) By $w_{\alpha\beta}$ we denote a directed link from node $\alpha$ to node $\beta$ with weight $w$. Because the significance clustering procedure described below works exactly the same for each particular state of the network, we omit the superscript of $G$ in what follows unless necessary to avoid confusion.

\subsubsection*{1. Cluster Real-World Network}
We first partition the network $G$ into the modular description $\mathsf{M}$. In the modular description, each node is assigned to one and only one module. The number of modules depends on the network and the objective function of the clustering algorithm. To capture the dynamics across the links and nodes in directed weighted networks, we use the map equation as the objective function \cite{theMapEquation2009,RosvallBergstrom08}. For a dynamic visualization of the mechanics of the map equation, see \url{http://www.mapequation.org}. In the supporting appendix, we present a short review and a new efficient algorithm to search for a partition of the network that minimizes the expected description length of a random walk across the nodes and links of the network. This description length is quantified by the map equation, but the search algorithm can also be generalized for other objective functions.

\subsubsection*{2. Generate and Cluster Bootstrap-World Networks}
The bootstrap is a statistical method for assessing the accuracy of an estimate by resampling from the empirical distribution. This method is particularly powerful when the variance of the estimator cannot be derived analytically or when the underlying distribution is not accessible. Because the cluster assignments are a result of a computational method and the network is idiosyncratic by nature, the bootstrap is indispensable for the process described here. 

To generate a single bootstrap replicate network $G_{b}^{*}$, we resample every link weight $w_{\alpha\beta}$ of the original network $G$ from a Poisson distribution with mean equal to the original link weight $w_{\alpha\beta}$. That is, $w_{\alpha\beta}^{*} \sim \text{Pois}(w_{\alpha\beta})$ for each link in the bootstrap network. Because of the parametric resampling of the link weights, this method formally falls under parametric bootstrapping. If the link weights cannot be modeled by a Poisson process, or if the links are unweighted, the Poisson resampling should be replaced by an appropriate alternative resampling procedure (see for example refs. \cite{karrer_newman,gfeller}).

Subsequently we partition the bootstrap replicate network with the same clustering method we used on the original network; this yields the bootstrap modular description $\mathsf{M}_{b}^{*}$. This procedure --- generating a bootstrap replicate network and clustering it into modules --- is repeated to generate a large number $B \sim 1000$ of bootstrap modular descriptions $\mathsf{M}^{*} = \{\mathsf{M}_{1}^{*}, \mathsf{M}_{2}^{*}, \ldots, \mathsf{M}_{B}^{*}\}$. Each {\em Bootstrap world} panel in Fig.~\ref{networktosankey_SI} illustrates four of these modular descriptions for four different bootstrap replicate networks, each created by the Poisson resampling procedure described above. Because this step requires that we cluster approximately 1000 bootstrap networks per time point, we have developed a new fast stochastic and recursive search algorithm for finding an accurate modular description of a given network (see the supporting appendix). 

\subsubsection*{3. Identify Significant Assignments}
The basic idea behind significance clustering is that we can look at the bootstrap replicates to see which aspects of the modular description of the original network are best supported by the data. Features of the original network that occur in all or nearly all of the bootstrap replicates are well-supported by the data; features that occur in only some of the bootstrap replicates are less well-supported.

What features do we consider? First, we consider the assignment of each node to a module.  By looking at the set of bootstrap modular descriptions, we can assess which of these assignments are strongly supported by the data, and which node assignments are less certain. To identify the nodes that are significantly assigned to a module, we search for the largest subset of nodes in each module of the original modular description $\mathsf{M}$ that are also clustered together in at least 95\% of all bootstrap modular descriptions $\mathsf{M}^{*}$. To pick the largest subset, of course we need some measure of size. The size of a subset could simply correspond to the number of nodes in the subset, but in line with our general clustering philosophy, we use the volume of flow through the subset. This is the total \emph{PageRank} of the cluster, which corresponds to the steady-state flow of random walkers that we use in the information-theoretic clustering algorithm.

To efficiently search the large space of possible subsets in each cluster, we use simulated annealing \cite{kirkpatrick}. Initially the nodes are randomly assigned to be members or non-members of the candidate largest subset. The score $S$ of the configuration is the size of the subset minus a penalty to account for the constraint that only nodes that are clustered together in at least 95\% of all bootstrap modular descriptions should be included. To implement the penalty, we first, and for each bootstrap modular description, count the number of nodes in the subset that do not belong to the largest group of nodes assigned to the same cluster. These are the mismatch nodes that break the constraint. To allow for a 5\% error, we add together the number of mismatch nodes for all bootstrap modular descriptions, excepting the 5\% with the highest number of mismatches. Finally we multiply this sum by ten times the cluster size, to make sure that the subset size and the penalty are of comparable size. This is necessary for an efficient search and a zero penalty at the end of the procedure (this {\em ad hoc} scaling factor of 10 was found by optimizing the convergence to a configuration with zero penalty and maximal subset size). After initiating with random assignments, we follow the standard simulated annealing scheme \cite{kirkpatrick}. At successively lower temperatures $T$, a node's subset assignment (member or non-member) is flipped and the score $S'$ for the new state is calculated. As in the Metropolis-Hastings algorithm \cite{metropolis1953,hastings1970}, the new state is always accepted if the new score is higher ($\Delta S = S' - S > 0$) or, conversely, if the new score is lower, the new state is accepted with probability equal to the Boltzmann factor of the score difference $\exp(\Delta S/T)$. Starting at $T=1$, we iterate this step as many times as there are nodes in the cluster, and then reduce the temperature according to $T' = 0.99T$. We repeat this procedure for as long as at least one new state is accepted for a given temperature. The nodes assigned to the subset at the final state serve as our approximation of the largest significant subset.

In addition to telling us about the assignment of individual nodes to specific modules, the set of bootstrap replicates also contains information about which modules stand alone and which are possibly subsets of other modules. To reveal this information, we need to identify the modules that are always, or almost always, separate from any other module.  We consider a module to be significant if its significant subset is clustered with no other significant subset in at least 95\% of all bootstrap modular descriptions. Conversely, two clusters are mutually nonsignificant if their significant subsets are clustered together in more than 5\% of all bootstrap modular descriptions.
In this way, each module can be mutually nonsignificant with more than one other module. In the alluvial diagram described in section 4, we want to associate each nonsignificant module with the module together with which it most likely forms a subset. The search for these pairs of modules is straightforward: For each pair of modules, we count in how many bootstrap modular descriptions all nodes in the two significant subsets are clustered together and record this number if it exceeds 5\% of all bootstrap modular descriptions (the criterion for nonsignificant modules). Then, starting at the smallest module, we associate the module with the other larger module with which it is most often clustered, and proceed to the next smallest module, and so on.

\subsubsection*{4. Construct Alluvial Diagram}
To reveal change over time or between states of real-world networks, we summarize the results of the significance clusterings of the different states $G^1, G^2, \ldots$ in an alluvial diagram. The diagram is constructed to highlight the significant changes, fusions, and fissions that the modules undergo between each pair of successive states $G^i$ and $G^{i+1}$. Each significance clustering for a state $G^i$ occupies a column in the diagram and is horizontally connected to preceding and succeeding significance clusterings by stream fields. Each block in a row of the alluvial diagram represents a cluster, and the height of the block reflects the size of the cluster (here in units of flow through the cluster, though other size measures, such as number of nodes, could be used instead). 
The clusters are ordered from bottom to top by size, with mutually nonsignificant clusters placed together and separated by a third of the standard spacing.
We use a darker color to indicate the significant subset of each cluster. Different colors can be used for clusters or groups of clusters to highlight particular stories in the data.

We use the stream fields to reveal the changes in cluster assignments and in level of significance between two adjacent significance clusterings. The height of a stream field at each end, going from the significant or nonsignificant subset of a cluster in one column to the significant or nonsignificant subset of a cluster in the adjacent column, represents the total size of the nodes that make this particular transition. By following all stream fields from a cluster to an adjacent column, it is therefore possible to  study in detail the mergers with other clusters and the significance transitions. To reduce the number of crossing stream fields, the stream fields are ordered by the position of the clusters to which they connect. For smooth transitions, we draw the stream fields with splines and use gradient shading for the component colors. Finally, to reduce visual clutter and improve clarity, we apply a threshold and do not show the thinnest stream fields.

\section*{Acknowledgments}
The authors would like to thank Jevin West both for processing the journal citation data and for numerous helpful discussions, and Moritz Stefaner for his extensive help with the information visualizations presented here.

% \bibliography{arxiv_timemap}
% \bibliographystyle{pnas}

\clearpage

\noindent \textbf{\large\sffamily Supporting appendix: Mapping directed weighted networks}\\

\noindent Here we briefly review our information theoretic approach to revealing community structure in weighted and directed networks \cite{RosvallBergstrom08sup} and present a new fast stochastic and recursive search algorithm to minimize the map equation --- the objective function of our method. We have developed this algorithm to be able to accurately partition the large number of bootstrap networks \cite{mapcode}. The search algorithm can also be generalized for other objective functions.

\subsection*{The map equation}

The objective of our flow-based and information-theoretic method known as the map equation is to find the structures within a network that are significant with respect to how information or resources flow through that network. For a detailed description of the map equation, see ref.~\cite{theMapEquation2009sup}. For a dynamic visualization of the mechanics of the map equation, see \url{http://www.mapequation.org}. The following is a short review. 

There is a duality between the problem of compressing a data set, and the problem of detecting and extracting significant patterns or structures within those data \cite{shannon,rissanen1978,grunwald}. We have developed the map equation approach to make use of this duality to detect community structure within directed and weighted networks that inherently are characterized by flow. For a given network partition $\mathsf{M}$, the map equation specifies the theoretical limit $L(\mathsf{M})$ of how concisely we can describe the trajectory of a random walker on the network. The underlying code structure of the map equation is designed such that the description can be compressed if the network has regions in which the random walker tends to stay for a long time. Therefore, with a random walk as a proxy for real flow, minimizing the map equation over all possible network partitions reveals important aspects of network structure with respect to the dynamics on the network.

To take advantage of the regional structure of the network, one index codebook and $m$ module codebooks, one for each module in the network, are used to describe the random walker's movements. The module codebooks have codewords for nodes within each module (and exit codes to leave the module), which are derived from the node visit/exit frequencies of the random walker. The index codebook has codewords for the modules, which are derived from the module switch rates of the random walker. Therefore, the average length of the code describing a step of the random walker is the average length of codewords from the index codebook and the module codebooks weighted by their rates of use. This is the map equation:
\begin{align}
L(\mathsf{M}) = q_{\curvearrowright} H(\mathcal{Q}) + \sum_{i=1}^{m}p_{\circlearrowright}^iH(\mathcal{P}^i).
\end{align}
The first term of this equation gives the average number of bits necessary to describe movement between modules, and the second term gives the average number of bits necessary to describe movement within modules. In the first term, $q_{\curvearrowright}$ is the probability that the random walker switches modules on any given step, and $H(\mathcal{Q})$ is the entropy of the module names, i.e.\ the frequency-weighted average length of codewords in the index codebook. In the second term, $H(\mathcal{P}^i)$ is the entropy of the within-module movements --- including an ``exit code'' to signify departure from module $i$, i.e.\ the frequency-weighted average length of codewords in module codebook $i$ --- and the weight $p_{\circlearrowright}^i$ is the fraction of within-module movements that occur in module $i$, plus the probability of exiting module $i$ such that $\sum_{i=1}^mp_{\circlearrowright}^i=1+q_{\curvearrowright}$.

To efficiently describe a random walk using a two-level code of this sort, the choice of partition $\mathsf{M}$ must reflect the patterns of flow within the network, with each module corresponding to a cluster of nodes in which a random walker spends a long period of time before departing for another module. To find the best such partition, we therefore seek to minimize the map equation over all possible partitions $\mathsf{M}$.

\subsection*{Fast stochastic and recursive search algorithm}

Any greedy (fast but inaccurate) or Monte Carlo-based (accurate but slow) approach can be used to minimize the map equation. To provide a good balance between the two extremes, we have developed a fast stochastic and recursive search algorithm, implemented it in C++, and made it available online both for directed and undirected weighted networks \cite{mapcode}. As a reference, the new algorithm is as fast as the previous high-speed algorithms (the greedy search presented in the supporting appendix of ref.~\cite{RosvallBergstrom08sup}), which were based on the method introduced in ref.~\cite{clauset-2004-70} and refined in ref.~\cite{wakita}. At the same time, it is also more accurate than our previous high-accuracy algorithm (a simulated annealing approach) presented in the same supporting appendix.

The core of the algorithm follows closely the method presented in ref.~\cite{blondel2008}: neighboring nodes are joined into modules, which subsequently are joined into supermodules and so on. First, each node is assigned to its own module. Then, in random sequential order, each node is moved to the neighboring module that results in the largest decrease of the map equation. If no move results in a decrease of the map equation, the node stays in its original module. This procedure is repeated, each time in a new random sequential order, until no move generates a decrease of the map equation. Now the network is rebuilt, with the modules of the last level forming the nodes at this level. And exactly as at the previous level, the nodes are joined into modules. This hierarchical rebuilding of the network is repeated until the map equation cannot be reduced further. Except for the random sequence order, this is the algorithm described in ref.~\cite{blondel2008}.

With this algorithm, a fairly good clustering of the network can be found in a very short time. Let us call this the core algorithm and see how it can be improved. The nodes assigned to the same module are forced to move jointly when the network is rebuilt. As a result, what was an optimal move early in the algorithm might have the opposite effect later in the algorithm. Because two or more modules that merge together and form one single module when the network is rebuilt can never be separated again in this algorithm, the accuracy can be improved by breaking the modules of the final state of the core algorithm in either of the two following ways:

\begin{itemize}
\item[] \emph{Submodule movements.} First, each cluster is treated as a network on its own and the main algorithm is applied to this network. This procedure generates one or more submodules for each module. Then all submodules are moved back to their respective modules of the previous step. At this stage, with the same partition as in the previous step but with each submodule being freely movable between the modules, the main algorithm is re-applied.

\item[] \emph{Single-node movements.} First, each node is re-assigned to be the sole member of its own module, in order to allow for single-node movements. Then all nodes are moved back to their respective modules of the previous step. At this stage, with the same partition as in the previous step but with each single node being freely movable between the modules, the main algorithm is re-applied.
\end{itemize}

In practice, we repeat the two extensions to the core algorithm in sequence and as long as the clustering is improved. Moreover, we apply the submodule movements recursively. That is, to find the submodules to be moved, the algorithm first splits the submodules into subsubmodules, subsubsubmodules, and so on until no further splits are possible. Finally, because the algorithm is stochastic and fast, we can restart the algorithm from scratch every time the clustering cannot be improved further and the algorithm stops. The implementation is straightforward and, by repeating the search 100 times, the final partition is less likely to correspond to a bad clustering of a local minimum. For each iteration, we record the clustering if the description length is shorter than the previously shortest description length. In practice, for the citation networks presented in this paper, which have on the order of 10,000 nodes and 1,000,000 directed and weighted links, each iteration takes about 5 seconds on a modern PC. We generate the significance clusterings by repeating the algorithm 100 times for each network and bootstrap network.\\

\end{document}